\documentclass{article}
\usepackage{frascatiphys}
\usepackage{graphicx}
\begin{document}
\title{ 
$P_c$ PHOTO-PRODUCTION AND DECAY
}
\author{
G.C.\ ROSSI\\
{\em  Dipartimento di Fisica, Universit\`a di  Roma
  ``{\it Tor Vergata}'' and INFN, Sezione di Roma 2} \\
  {\em Via della Ricerca Scientifica, 00133 Roma, Italy
    } \\
      {\em Centro Fermi - Museo Storico della Fisica e 
        Centro Studi e Ricerche ``E.\ Fermi'' } \\
        {\em Piazza del Viminale 1, 00184 Roma, Italy}\\\\
G.\ VENEZIANO\\
{\em  Coll\`ege de France, 11 place M. Berthelot, 75005 Paris, France} \\
{\em Theory Division, CERN, CH-1211 Geneva 23, Switzerland}
}
\maketitle
\baselineskip=11.6pt
\begin{abstract}
The 2015 LHCb discovery of a structure (denoted by $P_c^+$) decaying in $J/\psi \,p$ and conjectured to be a penta-quark state, has triggered a renewed interest in the question of possible existence of multi-quark states not predicted by the naive quark model. In this talk we present some considerations on $P_c$ photo-production experiments, aimed at testing its multi-quark interpretation in the framework of a 40-years-old ``string-junctionÕÕ picture that allows a unified description of baryons, tetra-, and penta-quark states.
\end{abstract}
\baselineskip=14pt

\section{Introduction and motivation}
\label{sec:INTRO}

In 2015 the LHCb collaboration\cite{Aaij:2015tga} announced the discovery of a massive ($M\simeq 4450$~MeV) and relatively narrow ($\Gamma \simeq 39$~MeV) baryonic state (denoted by $P^+_c$), decaying in the $J/\psi \,p$ channel, that does not fit in the naive quark model, which so nicely explains the spectrum of known mesons and baryons\cite{Tanabashi:2018oca}. 

The $P^+_c$ particle is interpreted as a state made by four quarks plus one antiquark (more precisely by $u\bar c cud$ quarks), just like ordinary mesons are composed by a quark and an anti-quark ($q \bar q$) and baryons by three quarks ($q q q$) arranged n a gauge invariant combination. Actually also (massive and narrow) states composed by two quarks and two antiquarks ($q q \bar q \bar q$) have been identified\cite{Abe:2003hq}\cite{Aaij:2013zoa}. A few reviews on the subject can be found in ref.\cite{REVIEWS}.
%Good reviews on the subject are\cite{Swanson:2006st}\cite{Bodwin:2013nua}\cite{Karliner:2017qhf}.
%\cite{Abe:2003hq,Aaij:2013zoa} \cite{Aaij:2017ueg}. denoted by $X(3872)$
The first hint to the existence of multi-quark states of the $q q \bar q \bar q$ kind can be found in the 1968 pioneering paper of ref.\cite{ROSNER} where they were conjectured as a way to alleviate problems with DHS duality\cite{DHS} in amplitudes involving also baryons. 

Experimentally the first claim of the discovery of a very narrow ($\Gamma\simeq 4$~MeV) and heavy $S(1940)$ state, coupled to the $p\bar p$ channel, dates back to 1974\cite{CAR}. This ``discovery'' triggered a lot of interest on the subject. The whole issue of multi-quark states was systematically taken up within the QCD framework in ref.\cite{Rossi:1977cy} where their existence was ``predicted'' on the basis of what we may call ``planar duality''. Indeed, in\cite{Rossi:1977cy}~\footnote{See also the review\cite{Montanet:1980te} and the more recent paper in ref.\cite{Rossi:2016szw}} it was proved that, as a solution of the duality constraints, emerging when besides meson-meson scattering amplitudes also meson-baryon and baryon-antibaryon amplitudes are considered, a new family of multi-quark states, endowed with an especially large coupling to baryons (thus dubbed ``baryonium'' states), must exist in the QCD spectrum. 

The peculiar coupling of these multi-quark states (baryonia) to baryons provides a possible natural explanation of the experimental observations that some of these states are very narrow (with the ratio width/mass less than 1/100). In ref.\cite{Rossi:1977cy} this feature was interpreted as due to the fact that the mass of (narrow) tetra-quark and penta-quarks states turn out to be smaller than the threshold for their favourite decay into fully baryonic channels (i.e.\ they have a mass smaller than two or, respectively, three times the baryon mass). Thus the decay into the dynamically favourite fully baryonic channels is kinematically forbidden and only the less favourite channels where only mesons or, respectively, one baryon plus mesons occur, are available. As we shall discuss, this scenario is supported by the string interpretation of hadron decay and formation in terms of color-string breaking and fusion.

Unfortunately the $S(1940)$ state was neither confirmed by experiments with much larger statistics\cite{AlstonGarnjost:1975vf} nor at LEAR, and people did forget about multi-quark states for nearly 30 years, until the striking discovery made in 2003 of the amazingly narrow ($\Gamma\simeq 3.5$~MeV) $X(3872)$ tetra-quark state by\cite{Abe:2003hq}, successively confirmed by\cite{Aaij:2013zoa}. Since then the zoo of multi-quark states has rapidly grown to tens of heavy (mostly narrow) states, though not all of them have survived more careful investigations and cross-checks and in the end entitled to an entry to the PDG heaven. Among the many states that did not survive more careful investigations we would like to mention the $\Theta_5$ penta-quark and the doubly charged $\Xi_{cc}^{++}$ baryonic state\cite{Aaij:2017ueg}, suggestively interpreted as penta-quarks belonging to an SU($N_f=3$)-decuplet. Actually at the moment even the existence of the $P^+_c$ penta-quark is at stake. Although it was confirmed by the analysis of ref.\cite{Aaij:2016phn}, it was not found in the very recent photo-production experiment carried out at JLAB\cite{Ali:2019lzf}. The message of this long historical overview is that multi-quark states are rather elusive. Convincingly establishing their existence requires a careful experimental and theoretical investigation.

In this talk we shall discuss the discovery potential of photo-production experiments in the search for the $P^+_c$ state in the theoretical framework of ref.\cite{Rossi:1977cy} where narrow tetra- and penta-quarks are interpreted as states of the baryonium family.

\section{The theoretical framework}
\label{sec:TF}

A unified picture of mesons, baryons and multi-quark states naturally emerges in the large $\lambda\equiv g^2N_c$ limit of QCD. Extending arguments first developed in\cite{Rossi:1977cy}, it was proved in\cite{Rossi:2016szw} that the strong coupling expansion of QCD is actually a large $\lambda$ expansion. In this limit meson and baryon propagators and their (four-point) scattering amplitudes are dominated by planar diagrams. Without entering into the details of the proof of this statement, we prefer to illustrate the situation with the help of a few figures. 
\vspace{-.2cm}

\subsection{Propagators}
\label{sec:PROP}

\begin{figure}[htb]
\begin{center}
{\includegraphics[height=0.38\linewidth]{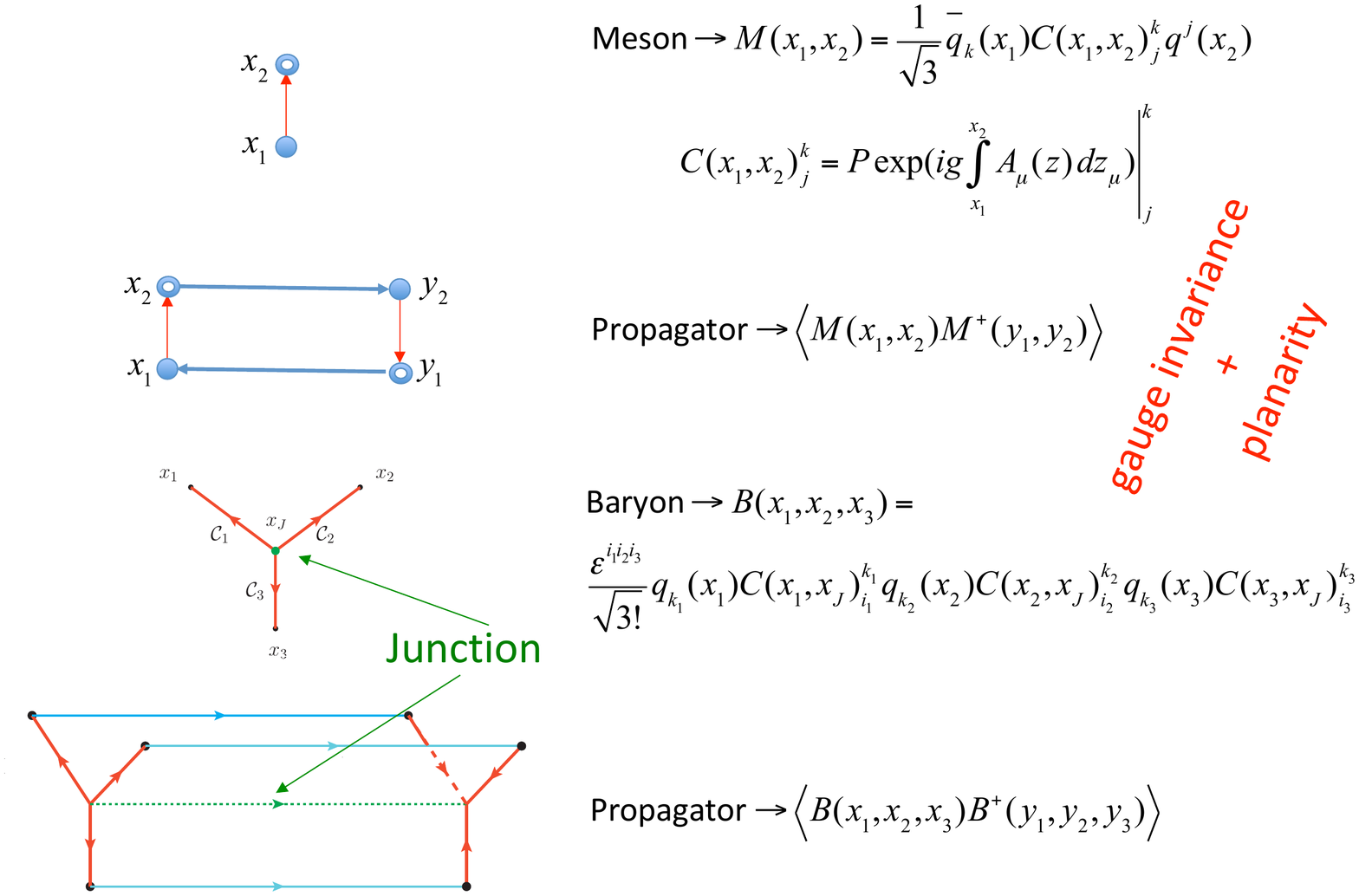}}
\vspace{-.4cm}
\caption{\it Meson and baryon propagators in the large $\lambda$ limit.}
\label{fig:fig1}
\end{center}
\vspace{-.3cm}
\end{figure}

We show in fig.\ref{fig:fig1} the gauge invariant expression of meson (two upper panels) and baryon (two lower panels) states and the world sheet spanned by their propagation. The blue lines represent the propagation of the quarks while the red segments are color flux lines. As well known in this way meson and baryons are described in terms of color strings attached to quarks and/or anti-quarks in a gauge invariant way.

A crucial consequence of the emerging topology of the diagrams describing the baryon propagation is the dynamical appearance of a special point (called junction) where the three ($N_c=3$) Wilson lines departing from the three quarks join to make up a gauge invariant operator. Thus in the large $\lambda$ approximation one can identify along the baryon propagator a line (drawn in green) that describes the color flow, or more physically the baryon number flow. 

This suggestive picture is beautifully confirmed by the existing lattice simulations\cite{Bissey:2006bz}\cite{Koma:2017hcm}. The simulations of ref.\cite{Bissey:2006bz} give strong support to the $Y$-shaped picture of the baryon depicted in fig.\ref{fig:fig1} showing that for sufficiently elongated strings color flux tubes where vacuum fluctuations are suppressed get formed that connect the three quarks in a $Y$-shaped arrangement. 

Further evidence for the $Y$-shaped picture of baryons comes from the more recent investigation reported in ref.\cite{Koma:2017hcm}. These authors are able to numerically determine the salient features of the three-quark potential showing that the latter is such that at each time the points belonging to the junction line lie at the minimal distance from the location of the three quarks, i.e.\ at the solution of the Fermat--Torricelli three-point minimal distance problem\cite{FT}. This is precisely what is predicted to occur by the ``bound-book'' structure of the diagrams describing the baryon propagation.
\vspace{-.2cm}

\subsection{Amplitudes}
\label{sec:AMPL}

\begin{figure}[htb]
    \begin{center}
        {\includegraphics[height=0.22\linewidth]{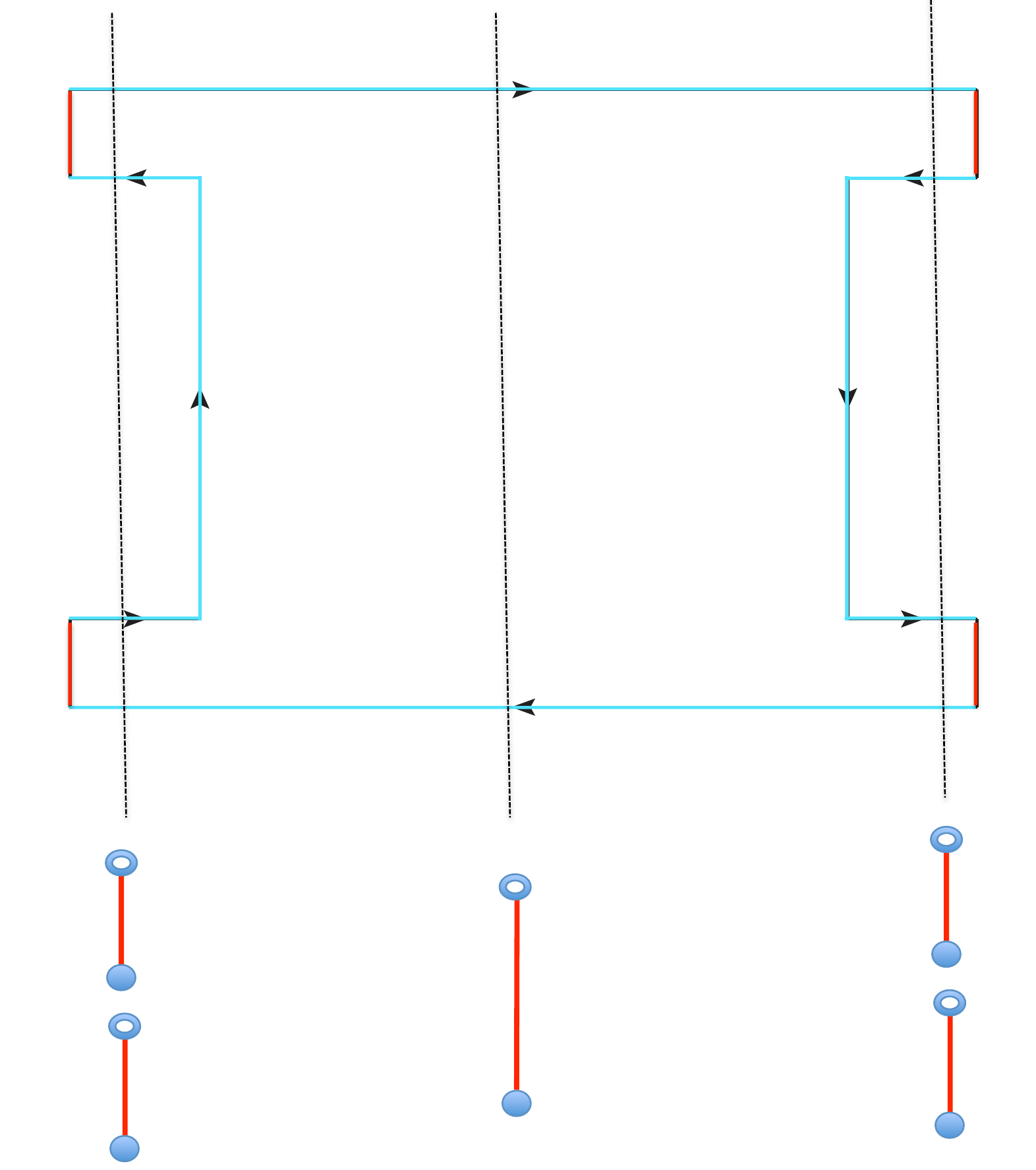}}\hspace{2.cm}
        {\includegraphics[height=0.20\linewidth]{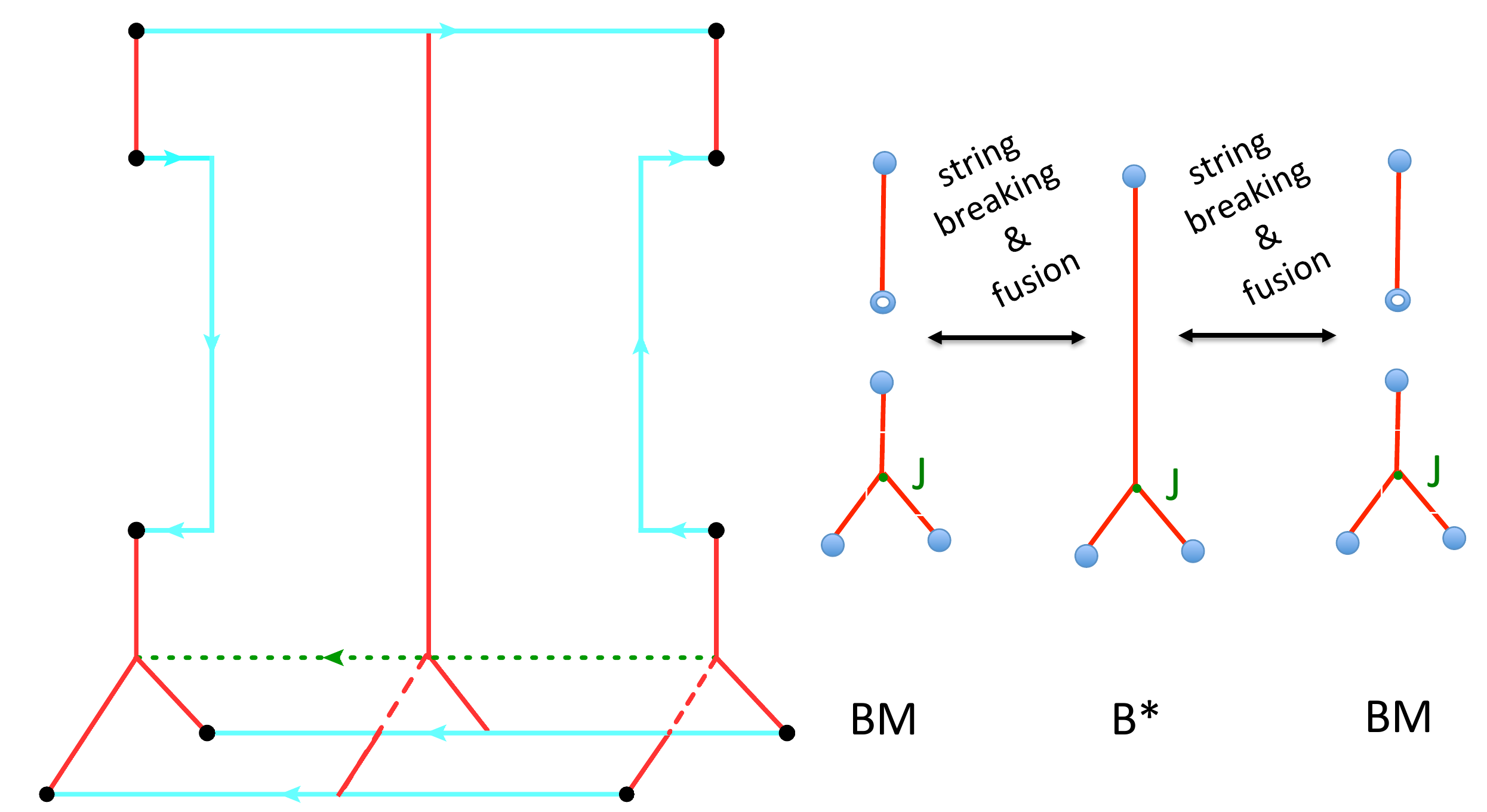}}
        \caption{\it Left: $MM\to MM$ scattering amplitude - Right: $MB\to MB$ scattering amplitude.}
\label{fig:fig2}
    \end{center}
    \vspace{-.4cm}
\end{figure}

Glueing together meson strips and appropriately stretching the resulting surface one immediately gets the planar topology describing the $MM\to MM$ scattering amplitude to leading order in $\lambda$. This is shown in the left panel of fig.\ref{fig:fig2}. key observation about this figure is that meson-meson fusion and meson decay occur through color string fusion and breaking process, respectively. 

Glueing meson to baryon sheets leads to the $MB\to MB$ amplitude depicted in the right panel of fig.\ref{fig:fig2} where we also display the string fusion and breaking processes taking place in the $s$-channel. 

\begin{figure}[htb]
    \begin{center}
        {\includegraphics[height=0.30\linewidth]{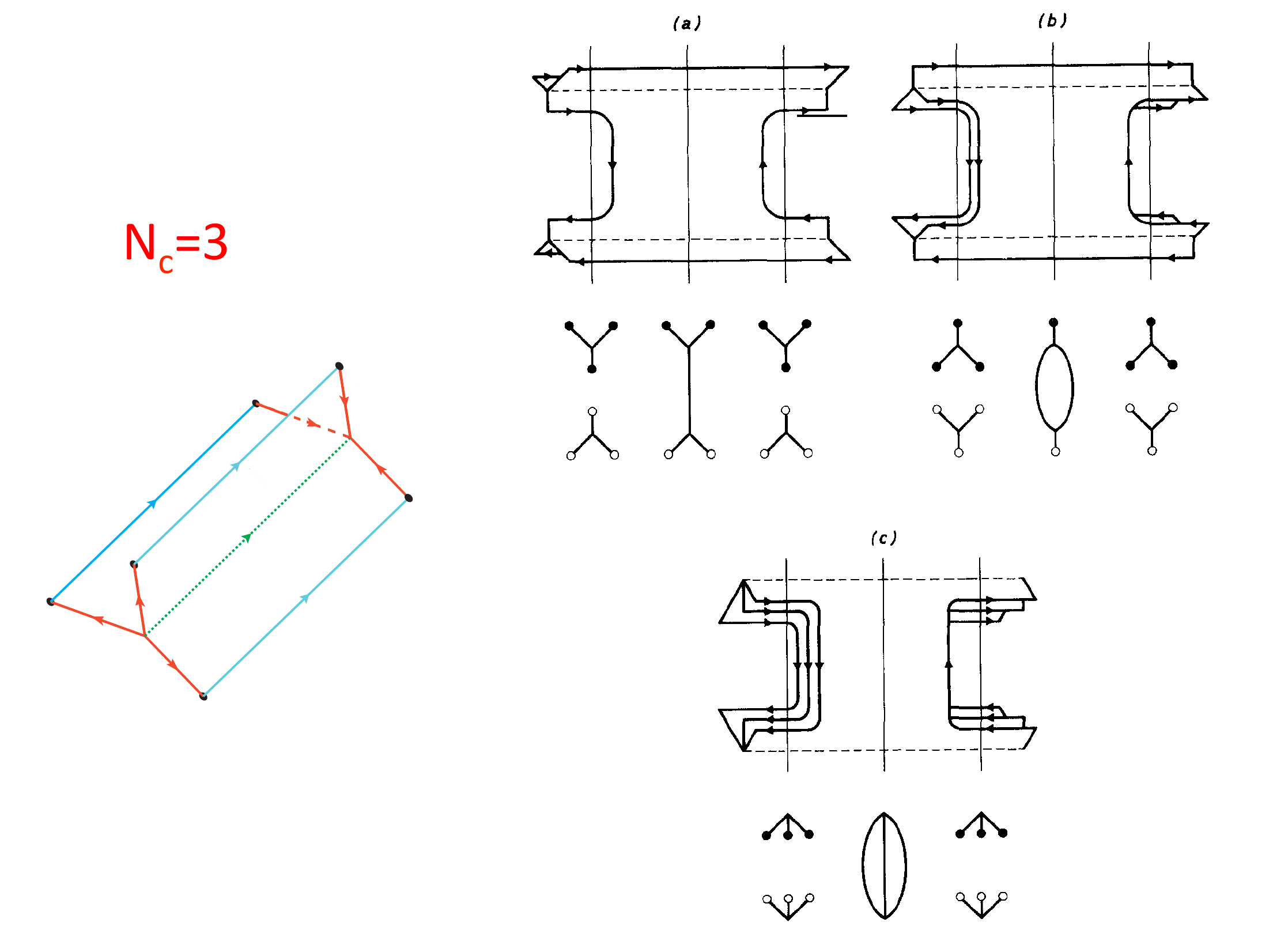}}\hspace{1.cm}
       {\includegraphics[height=0.18\linewidth]{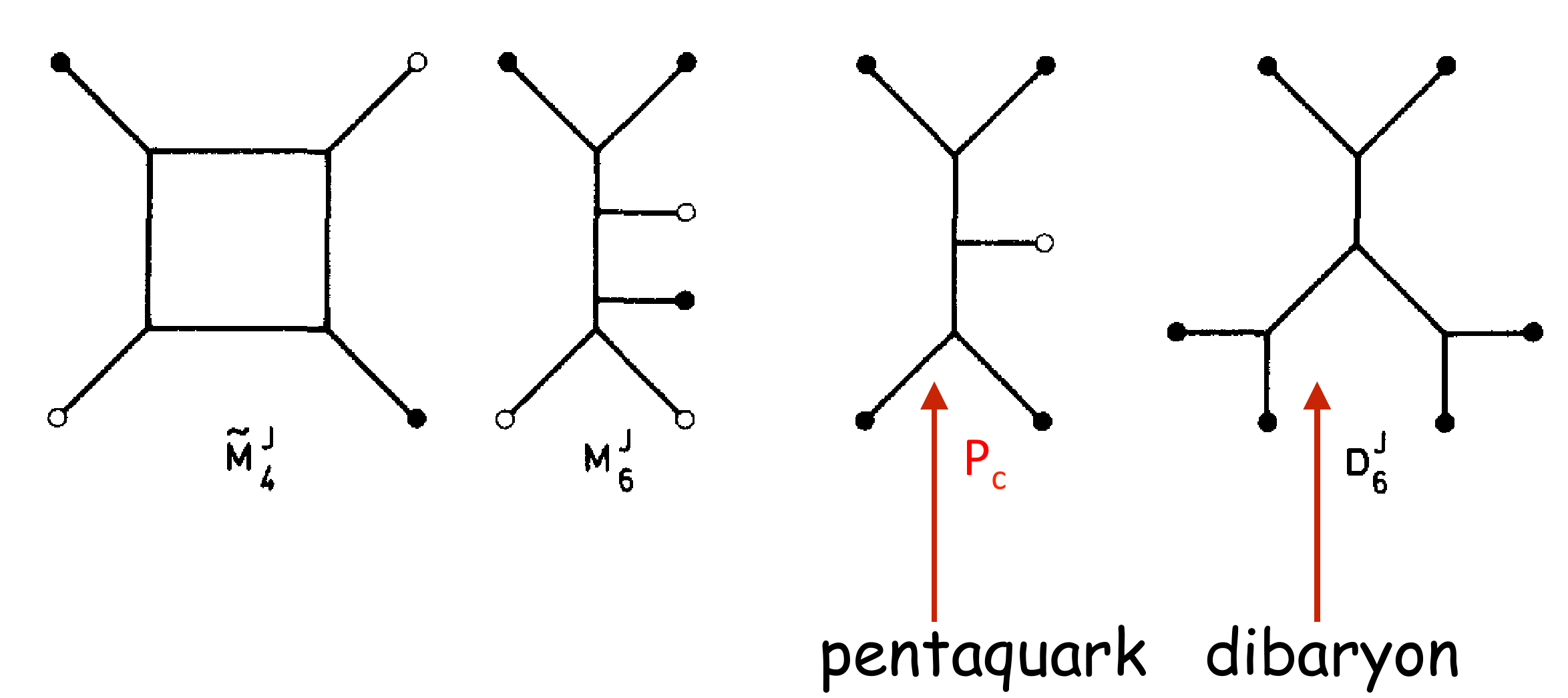}}
       \vspace{-.2cm}
 \caption{\it Left: $B\bar B \to B\bar B$ scattering amplitudes - Right: Members of the baryonium family.}
\label{fig:fig3}
    \end{center}
    \vspace{-.4cm}
\end{figure}

$B\bar B\to B\bar B$ amplitudes are constructed by glueing the sheets of the two ``bound-books'' representing the propagating baryons. In doing that we have a number of possibilities (three, if $N_c=3$), shown in the right panel of fig.\ref{fig:fig3}. New $s$-channel states endowed with a junction ($J$) and an anti-junction ($\bar J$) are formed. Among them we find states made by two quarks and two anti-quarks that we shall denote $M_4^J$. 

Notice that besides the scattering amplitudes displayed in the figure one should add three more types of amplitudes describing annihilation. They are simply obtained from the previous diagrams by a $90^\circ$ rotation. The difference between the two sets of diagrams is that in the former the $J$ and $\bar J$ lines ($B$ and $\bar B$ baryon number) flow undisturbed from the initial to the final state, while in the latter the $J$ and $\bar J$ lines annihilate in the initial state, giving raise to (jets of) ordinary mesons as intermediate $s$-channel states, from which a $J$ and $\bar J$ pair (i.e.\ a $B\bar B$ pair) is finally created. Thus following the flux of flavour is not enough to identity the structure of a diagram: one also has to specify the fate of the junctions.
\begin{figure}[htb]
\begin{center}
{\includegraphics[height=0.2\linewidth]{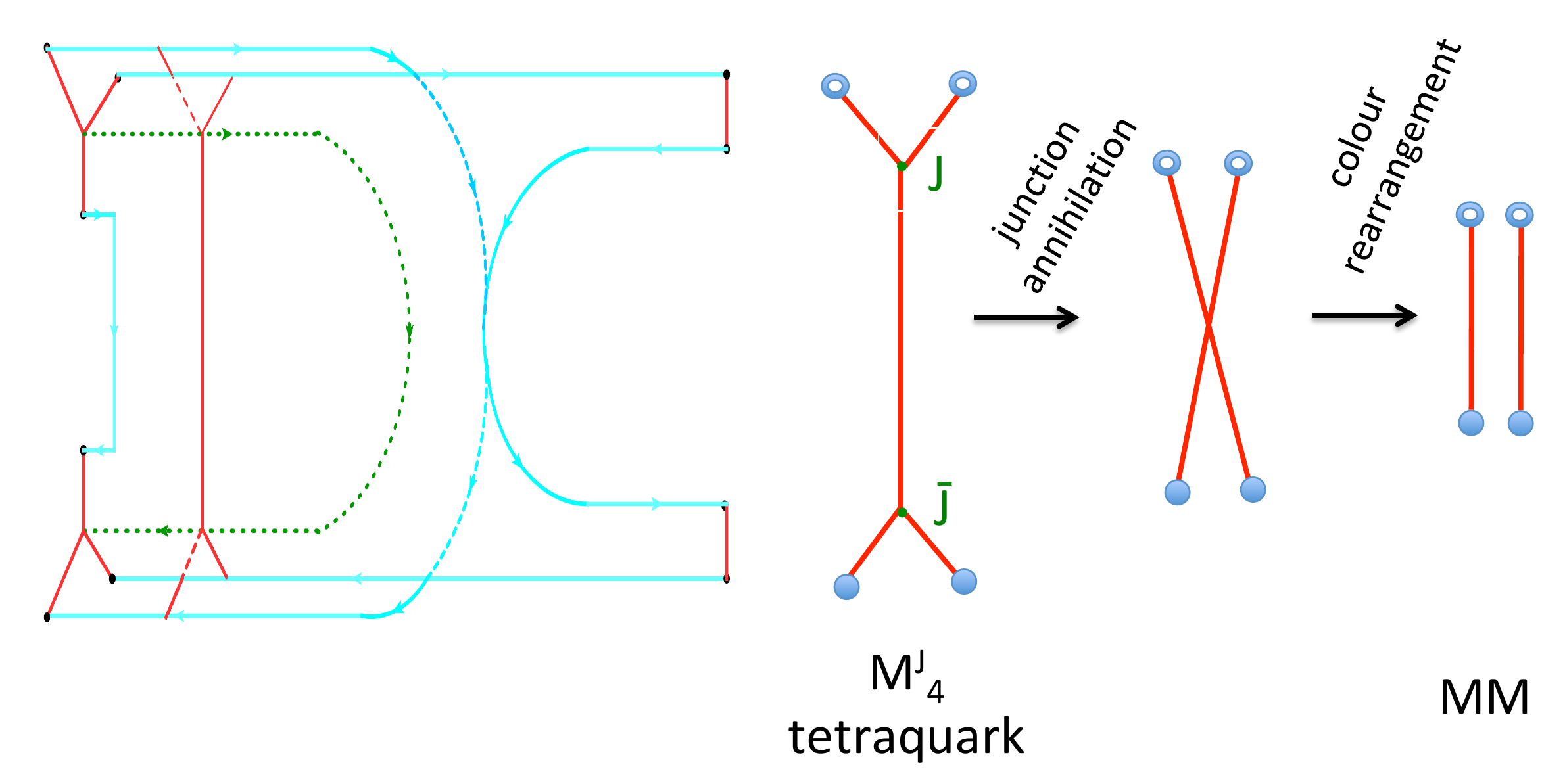}}
\caption{\it The non-planar $B\bar B\to MM$ annihilation amplitude.}
\label{fig:fig4}
\end{center}
\vspace{-.4cm}
\end{figure}

From more complicated amplitudes more complicated hadrons endowed with junctions and/or anti-junctions emerge as possible intermediate states. They can be easily constructed based on the principle of gauge invariance. A few of them are shown in the left panel of fig.\ref{fig:fig3}. Members of the baryonium family like dibaryons and penta-quarks were for the first time conjectured to exist in\cite{Montanet:1980te}.

%\subsection{Non-planar topologies}\label{sec:NPT}

We conclude by comparing the planar diagrams in the right panel of fig.\ref{fig:fig2} with the non-planar $B\bar B\to MM$ annihilation depicted in fig.\ref{fig:fig4}. We see that non-planar topologies entail a new kind of processes in which $M_4^J$ decays in two mesons via a $J-\bar J$ annihilation followed by ``colour rearrangement''.

\section{$P_c$ and photo-production}
\label{sec:PCPP}

\begin{figure}[htb]
\begin{center}
{\includegraphics[height=0.25\linewidth]{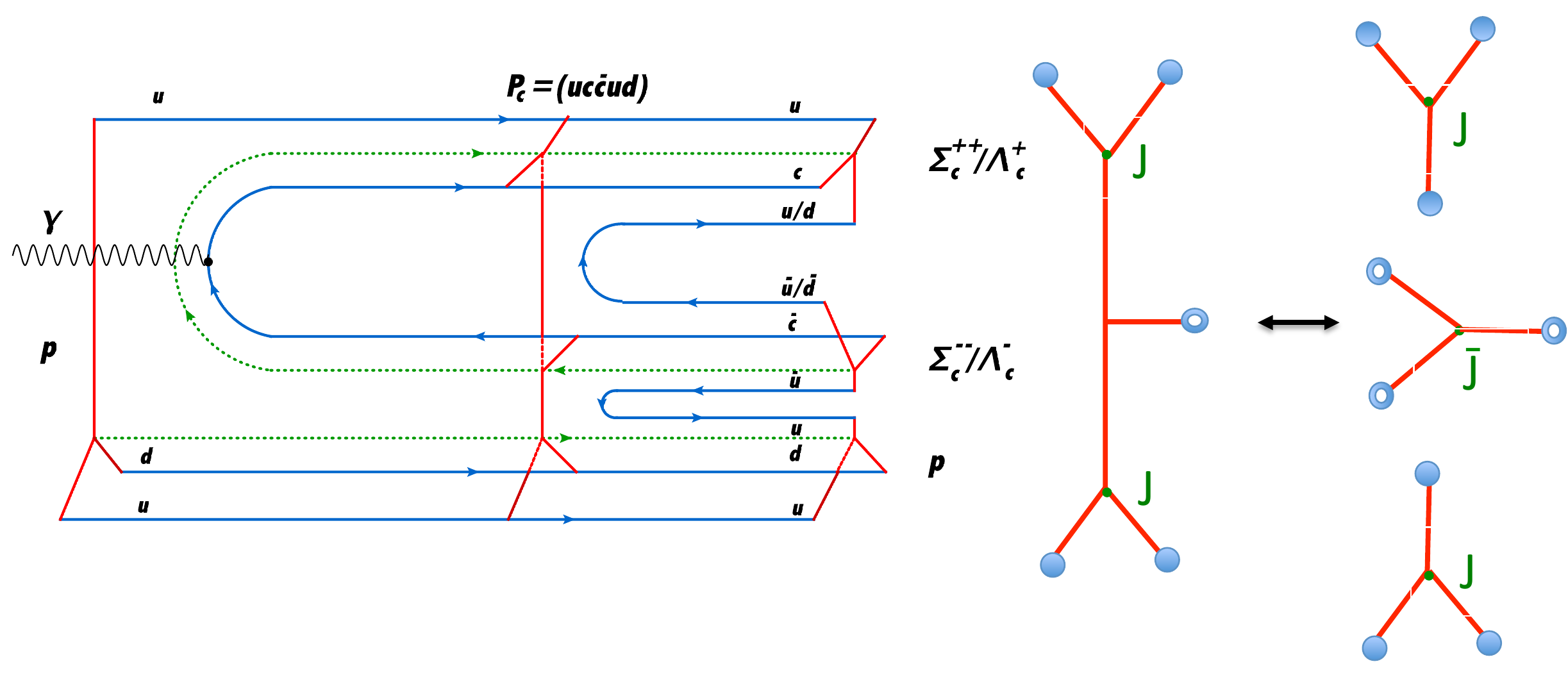}}
\vspace{-.2cm}
\caption{\it The kinematically forbidden $P_c\to B\bar B B$ decay.}
\label{fig:fig5}
\end{center}
\vspace{-.4cm}
\end{figure}

\begin{figure}[htb]
    \begin{center}
        {\includegraphics[height=0.25\linewidth]{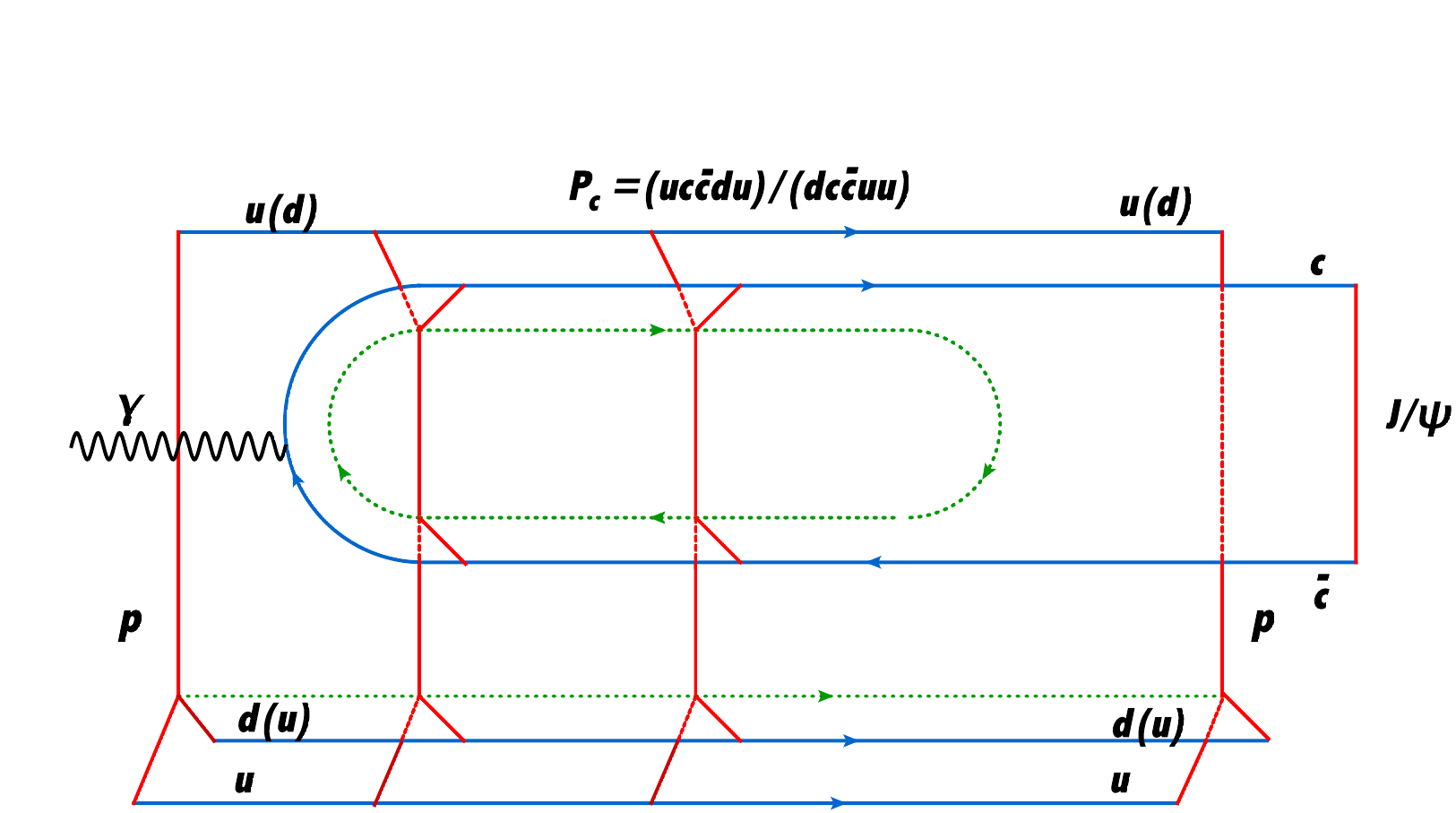}}\hspace{1.cm}
       {\includegraphics[height=0.2\linewidth]{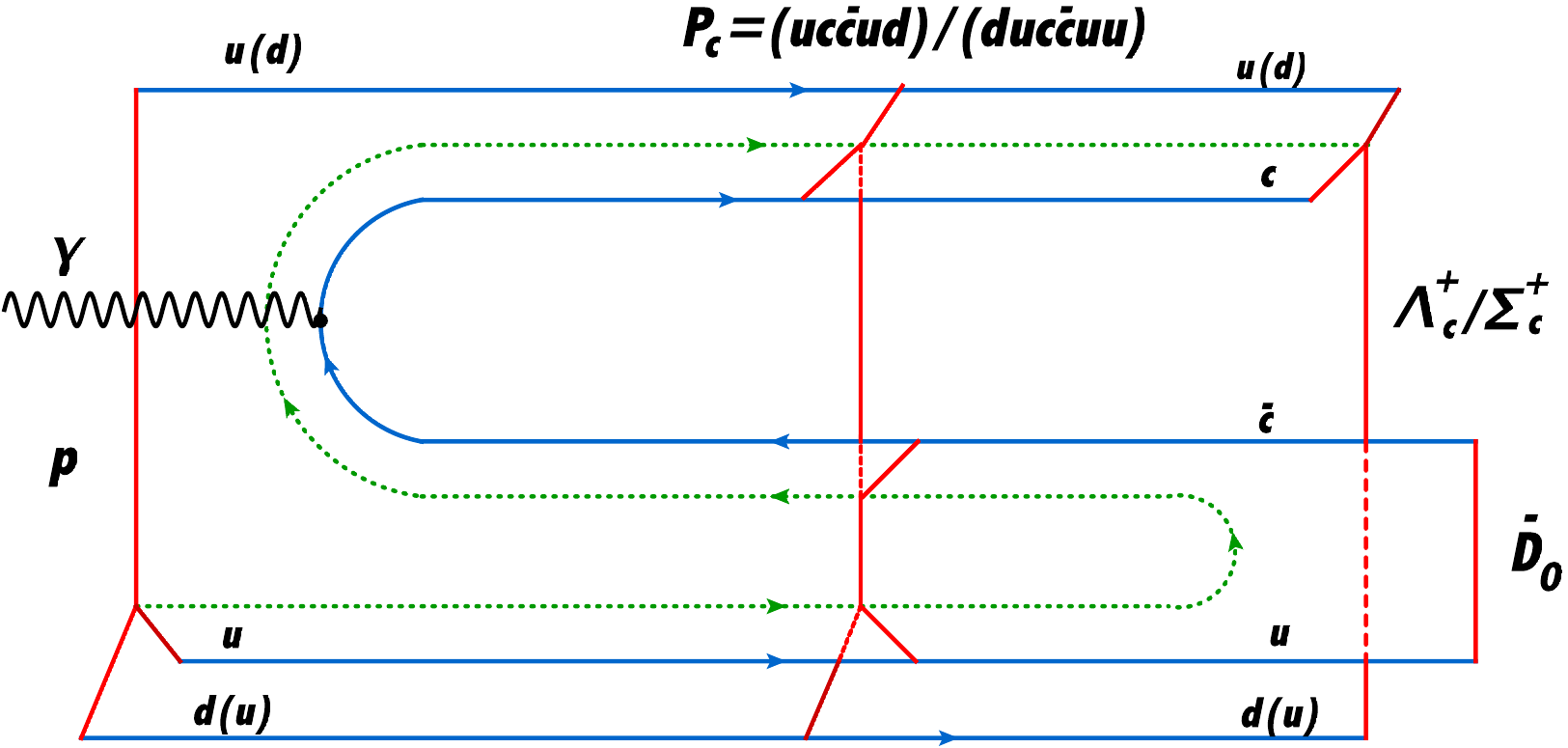}}
 \caption{\it Colour rearrangement $P_c$ decay. Left: $P_c\to J/\psi + p$ - Right: $P_c\to \Lambda_c^+ /\Sigma^+_c + \bar D_0$.}
\label{fig:fig6}
    \end{center}
    \vspace{-.4cm}
\end{figure}

$P_c[u\bar c cud]$ is a hidden charm state. So it is natural to think to photo-production on a proton target as an efficient way to create a $c\bar c$ pair out of the vacuum.
%\begin{figure}[htb]
%\begin{center}
%{\includegraphics[height=0.5\linewidth]{penta_LHCb.pdf}}
%\caption{\it The experimental $P_c$ evidence from\cite{Aaij:2015tga}.}
%\label{fig:fig4}
%\end{center}
%\end{figure}
If the $P_c$ mass was larger than the three-baryon threshold\footnote{The fully baryonic decays $P_c\to \Lambda_c^{+}+\Lambda_c^{-}+p$ and $P_c\to \Sigma_c^{++}+\Sigma_c^{--}+p$ are kinematically forbidden, since $M_{P_c}\!\sim\!  4450$ while $2 M_{\Lambda_c^{+}}\!+\!M_p\!\sim \!2\!\times\! 2286 \!+ \!938\!=\! 5510$~MeV and $2 M_{\Sigma_c^{++}}\!+\!M_p\!\sim\!2\times\! 2454 \!+ \!938\!= \!5846$~MeV.} the typical dynamically dominant (i.e.\ ``planar'') diagram would be the one shown in fig.\ref{fig:fig5} where the intermediate $P_c$ resonance decay proceeds via two string breakings.
This string breaking process is, however, kinematically forbidden and the decay needs to proceed via color rearrangement. Examples of such decays are $P_c\to J/\psi + p$, $P_c\to \Lambda_c^+ /\Sigma^+_c + \bar D_0$ and $P_c\to \Sigma^{++}_c + D^-$. The first two are depicted in fig.\ref{fig:fig6}. Among the three, both by phase space considerations and because of the possible dominance of diagrams where junctions are closed in a loop (``bathtub'' diagrams) over diagrams where the junction line unrolls through the diagram (``snake' diagrams), we think that the $P_c\to J/\psi + p$ decay drawn in the left panel of fig.\ref{fig:fig6} is the dominant process.
\vspace{-.4cm}
 %\subsection{Tagged $P_c$ photo-production}\label{sec:TRPCPP}
%If sufficiently high energy photons are available one can investigate the pattern of flavour partners of $P_c[u\bar c cud]$ with the help of a tagged production process in which also a pseudoscalar meson ($\pi$, $K$, $D$) is identified as shown in fig.\ref{fig:fig7}.

\begin{figure}[htb]
    \begin{center}
        {\includegraphics[height=0.2\linewidth]{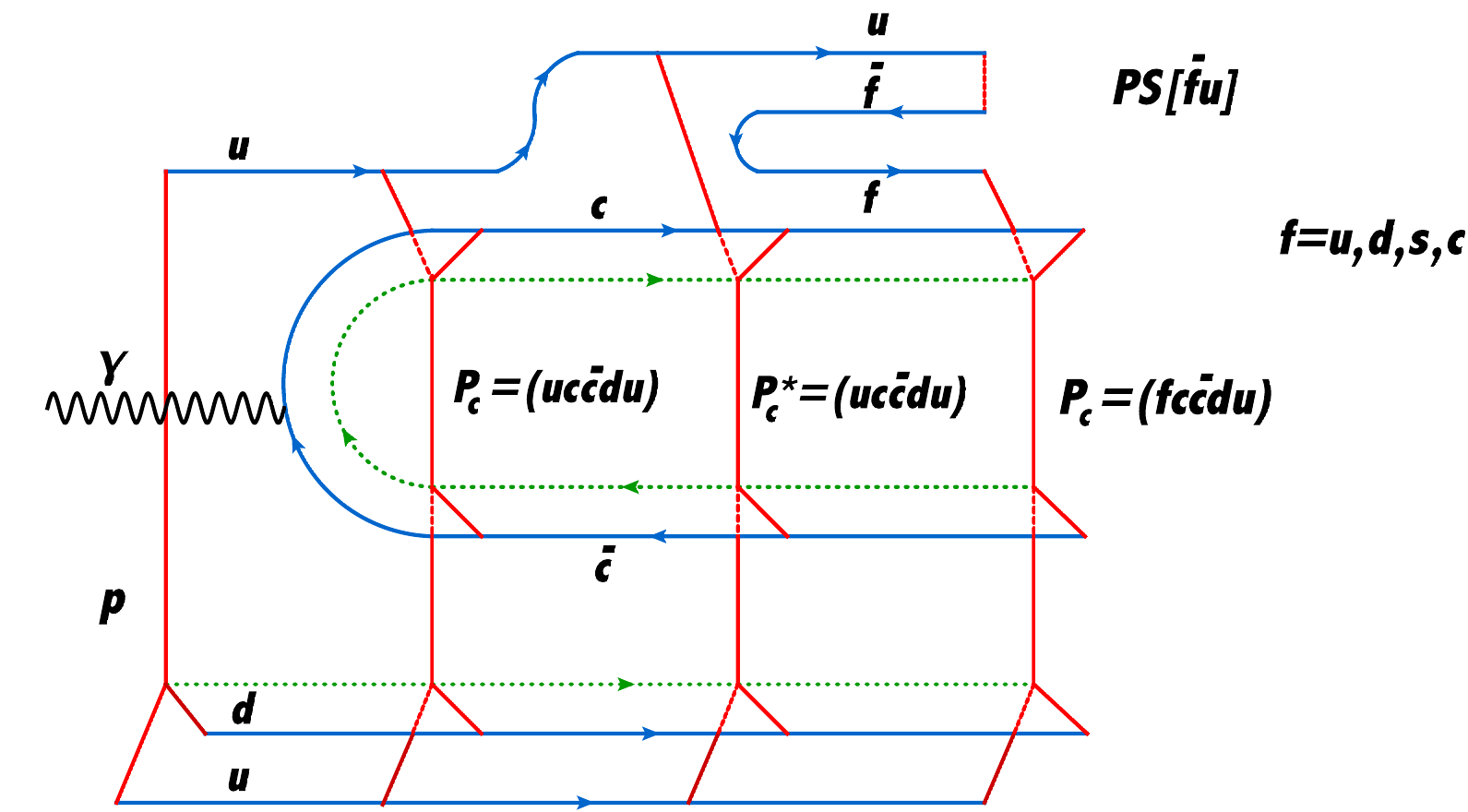}}\hspace{1.cm}
       {\includegraphics[height=0.2\linewidth]{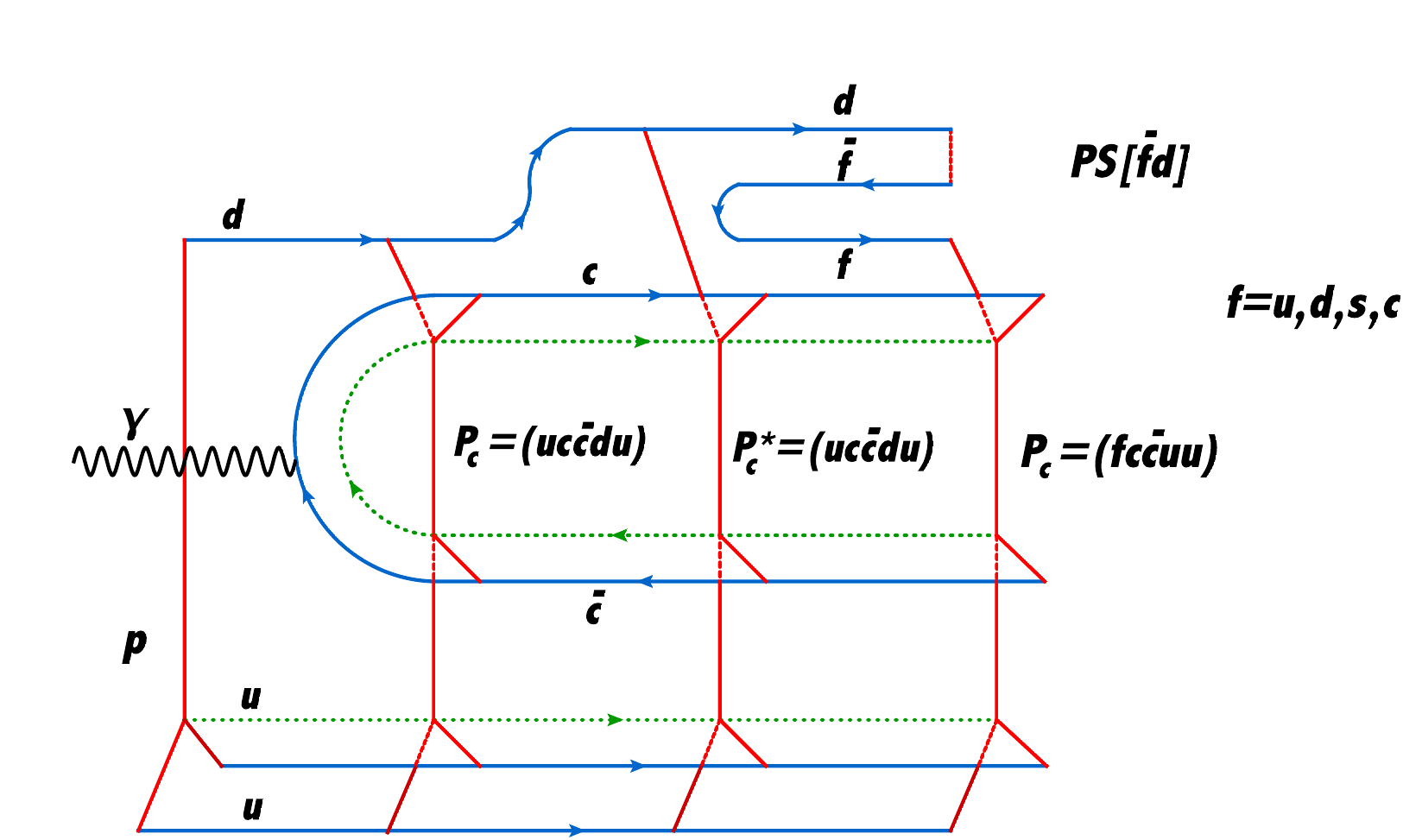}}
% \caption{\it Tagged photo-production. Left panel: $\gamma \, p\to P_c^\star (u \,c \,\bar c\, u\,d)\to  K^+ \,P_c^s(s \,c \,\bar c\, u\,d)$ - Right panel: .$\gamma \, p\to P_c^\star (d \,c \,\bar c\, u\,u)\to  K^0 \,P_c^s(s \,c \,\bar c\, u\,u)$}
 \caption{\it Left: $\gamma p\to P_c^\star [f c \bar cud]\to  PS[\bar f u] + P_c[f c \bar c ud]$ - Right: $\gamma p\to P_c^\star [f c \bar c u u]\to PS[\bar f d] +\,P_c[f c \bar c uu]$.}
\label{fig:fig7}
    \end{center}
    \vspace{-.4cm}
\end{figure}

If sufficiently energetic photons are available, the pattern of $P_c[u\bar c cud]$ flavour partners can be studied with the help of a tagged production process where a PS meson ($\pi$, $K$, $D$) is identified, see fig.\ref{fig:fig7}.
\vspace{-.4cm}
%\section{Acknowledgements} We wish to thank the Organizers for the warm and exciting atmosphere we found in  the Conference.

\subsection{Peak resolution}
\label{sec:PRES}

The recent resolution of the $P_c$ peaks\cite{Aaij:2019vzc} indicates the presence of three resonances that have a similar interpretation in our picture and in the diquark model of ref.\cite{Maiani:2015vwa} (for an alternative scheme see\cite{Karliner:2015ina}). In our scenario the lighter peak (4312~MeV) should be identified with the $(ud)_{I=0}\, \bar{c}\, (uc)$ state, while the two heavier and almost degenerate ones (4450 \& 4457~MeV) with $(ud)_{I=1} \,\bar{c} \,(uc)$ \& $(uu)_{I=1}\, \bar{c} \,(dc)$ (see fig.\ref{fig:fig6}).

\end{document}